\documentstyle{article}
\input math_macros
\font\tenbf=cmbx10
\font\tenrm=cmr10
\font\tenit=cmti10
\font\elevenbf=cmbx10 scaled\magstep 1
\font\elevenrm=cmr10 scaled\magstep 1
\font\elevenit=cmti10 scaled\magstep 1

\renewenvironment{thebibliography}[1]
 { \elevenrm
   \begin{list}{\arabic{enumi}.}
    {\usecounter{enumi} \setlength{\parsep}{0pt}
      \setlength{\itemsep}{3pt} \settowidth{\labelwidth}{#1.}
       \sloppy
      }}{\end{list}}

\begin{document}
\rightline{PURD-TH-92-14}
\begin{center}{{\tenbf QUARK-LEPTON SYMMETRY}
\vglue 5pt
\vglue 0.5cm
{\tenrm
R. FOOT$^{(a)}$, H. LEW$^{(b)}$\footnote{Talk given at the
DPF92 meeting, Fermilab, November 1992.}
and R. R. VOLKAS$^{(c)}$\\}
\vglue 0.5cm
\baselineskip=13pt
{\tenit $^{(a)}$ Physics Department, McGill University,
3600 University St.,\\}
\baselineskip=12pt
{\tenit Montr\'eal, Qu\'ebec, Canada H3A 2T8.\\}
\vglue 0.3cm
{\tenit $^{(b)}$ Physics Department, Purdue University,\\}
\baselineskip=12pt
{\tenit West Lafayette, IN 47907-1396, U.S.A.\\}
\vglue 0.3cm
{\tenit $^{(c)}$ Research Centre for High Energy Physics, School of Physics,
University of Melbourne,\\}
\baselineskip=12pt
{\tenit Parkville 3052, Australia\\}
\vglue 0.5cm
{\tenrm ABSTRACT}}
\end{center}
\vglue 0.1cm
{\rightskip=3pc
 \leftskip=3pc
 \tenrm\baselineskip=12pt
 \noindent
Quark-lepton symmetric models are a class of gauge theories
motivated by the similarities between the quarks and leptons.
In these models the gauge group of the standard model is
extended to include a ``color'' group for the leptons.
Consequently, the quarks and leptons can then be related by a
$Z_2$ discrete quark-lepton symmetry which is spontaneously broken
by the vacuum. Models utilizing quark-lepton symmetry with acceptable
and interesting collider phenomenology have been constructed. The
cosmological consequences of these models are also discussed.
\vglue 0.6cm}

{\elevenbf\noindent 1. The minimal quark-lepton symmetric model} \cite{flql}
\vglue 0.2cm
\baselineskip=14pt
\elevenrm
Historically, there has been a rough correspondence between
the hadrons and leptons which is now translated into a
quark-lepton ``symmetry''. This rough correspondence can be seen
in the similarities of the weak interactions for the quarks
and leptons. Therefore it maybe interesting to postulate an
{\elevenit exact } symmetry between quarks and leptons in the context
of gauge theories.
In the Minimal Standard Model (MSM), the quarks carry color
whereas the leptons do not and it is also assumed that there
are no right-handed neutrinos. To implement quark-lepton
symmetry (hereafter referred to as q-$\ell$ symmetry),
equal numbers of quark and lepton degrees of freedom are needed.
To achieve this we will introduce (i) the right-handed neutrino,
$\nu_R$, and (ii) a ``color'' group for the leptons. This then
necessitates extending the MSM gauge group, $G_{SM}$, to
$\ G_{q\ell} = SU(3)_{\ell} \otimes SU(3)_q \otimes SU(2)_L \otimes
U(1)_X \ $
supplemented by a $Z_2$ discrete symmetry between the quarks
and leptons. Here SU(3)$_q$ is the usual color group and
SU(3)$_{\ell}$ is its leptonic partner.
The expanded fermionic generation is defined by the transformation laws
\begin{eqnarray}
Q_L &\sim &(1,3,2)(1/3),\ \ u_R \sim (1,3,1)(4/3),\ \ d_R \sim
(1,3,1)(-2/3), \nonumber \\
F_L &\sim & (3,1,2)(-1/3),\ \ E_R \sim (3,1,1)(-4/3),\ \ N_R \sim
(3,1,1)(2/3). \nonumber
\end{eqnarray}
The $Z_2$ discrete symmetry:
$F_L \leftrightarrow Q_L,\ E_R \leftrightarrow u_R,\ N_R
\leftrightarrow d_R,\
G^{\mu}_q \leftrightarrow G^{\mu}_{\ell},\ C^{\mu}
\leftrightarrow -C^{\mu}$
can now be defined [where $G^{\mu}_{q,\ell}$ are the gauge bosons
of SU(3)$_{q,\ell}$ and $C^{\mu}$
is the gauge boson of U(1)$_X$].
Standard hypercharge is given by
$Y = X + {1\over 3}T$, where
$T = {\rm diag}(-2,1,1)$ is a generator of SU(3)$_{\ell}$. Standard
leptons are identified with the $T=-2$ components of the leptonic color
triplets, while the $T=1$ components are the exotic charge $\pm$1/2
leptons, called liptons.

The MSM works very well for energies up to about 100 GeV.
In order to spontaneously break SU(3)$_{\ell}$ and the
quark-lepton discrete symmetry, as well as giving
mass to the liptons, the Higgs
bosons $\chi_1 \sim (\overline{3},1,1)(-2/3)$ and
$\chi_2 \sim (1,\overline{3},1)(2/3)$ are introduced with
$\chi_1 \leftrightarrow \chi_2$.
The $T = 2$ component of $\chi_1$ develops a nonzero vacuum expectation
value (VEV), while the VEV of $\chi_2$ is completely zero.
Electroweak symmetry breaking is achieved through the Higgs
doublet, $\phi \sim (1,1,2)(1)$, with
$\phi \leftrightarrow \phi^c$(charge conjugate field)
under q-$\ell$ symmetry.
The overall symmetry breaking pattern can be summarised as follows:
\footnote{To show that the required pattern of symmetry breaking
for the minimal q-$\ell$ model can be realised, consider the
Higgs potential
$$ V = \lambda_1
\left[ \chi_1^\dagger \chi_1 + \chi_2^\dagger \chi_2 -v^2 \right]^2
 + \lambda_3 \left( \phi^\dagger \phi - u^2 \right)^2
 +  \lambda_2 \chi_1^\dagger \chi_1 \chi_2^\dagger \chi_2
 +  \lambda_4 \left[ \phi^\dagger \phi - u^2 + \chi_1^\dagger \chi_1
+ \chi_2^\dagger \chi_2 -v^2 \right]^2. $$}
$\ G_{ql} \buildrel {\langle\chi_1 \rangle} \over \longrightarrow
SU(2)' \otimes G_{SM}
\buildrel {\langle\phi\rangle} \over \longrightarrow
SU(2)' \otimes SU(3)_q \otimes U(1)_Q $.

This minimal Higgs sector results in the tree-level mass relations,
$M_u = M_e$ and $M_d =M^{\rm Dirac}_{\nu}$,
where $M_{u,e,d,\nu}$ refer to the $3 \times 3$ fermion mass matrices.
These mass relations arise as
a consequence of (i) the assumption that q-$\ell$ symmetry is
a symmetry of the Yukawa Lagrangian and (ii) using
only one Higgs doublet.\footnote{
The mass relation involving the neutrinos can be avoided if
Majorana masses are given to the right-handed neutrinos.
This can be achieved through the Higgs multiplets
$\Delta_1 \sim (6,1,1)(4/3)$ and $\Delta_2 \sim (1,6,1)(-4/3)$ with
$\Delta_1 \leftrightarrow \Delta_2$.
It is assumed that the $T=-4$ component of $\Delta_1$
develops a nonzero VEV while the VEV of $\Delta_2$ remains zero.}
If the minimal model is extended to
contain two Higgs doublets, then the abovementioned mass relations
can be avoided at tree-level but at the expense of predictivity.
Alternatively, a certain q-$\ell$ symmetric
model with a non-minimal gauge group has been
shown to contain radiative corrections which can yield correct but
unpredictive fermion masses\cite{flmass}.
\vglue 0.6cm
{\elevenbf\noindent 2. Phenomenological Implications} \cite{flvph}
\vglue 0.2cm
{\elevenit\noindent 2.1. Gauge Bosons}
\vglue 0.1cm
\baselineskip=14pt
\elevenrm
The low energy gauge group of the q-$\ell$ model is
$SU(2)' \otimes SU(3)_q \otimes U(1)_Q$. As a result of the
larger symmetry there will be additional gauge bosons, both massive
and massless, to that of the MSM. The decomposition of the
$SU(3)_\ell$ gauge bosons under the low energy gauge group is:
$\ 8 \rightarrow  (1,1)(0) \oplus (2,1)(-{1\over 2})
\oplus (2,1)({1\over 2}) \oplus (3,1)(0) \ $.
The neutral component, the $Z'$ boson, can mix with the standard Z boson.
The $SU(2)'$ doublets of charge $\pm {1\over 2}$ are the massive
$SU(3)_\ell / SU(2)'$ bosons which can contribute to FCNC processes
at 1-loop (e.g. $\mu \rightarrow e + \gamma$).\footnote{
This will put a bound on the mass of the coset space gauge bosons in
terms of mixing angles and lipton masses. There will also be contributions
from the Higgs sector which we have not considered.}
The remaining triplet contains the massless $SU(2)'$ gauge bosons.
By fitting the model parameters
to the low energy electroweak data, the lower bound on
the mass of the $Z'$ boson was found to be around 700 GeV.
If the number of fermions is not too large then $SU(2)'$ is
expected to be asymptotically free and by analogy with QCD
we assume that it confines all SU(2)$'$ colored states.
Thus we expect $SU(2)'$ glueballs
to exist with a mass of the order of the $SU(2)'$ confinement
scale, $\Lambda_{SU(2)'}$ .
\vglue 0.2cm
{\elevenit\noindent 2.2. Liptons}
\vglue 0.1cm
The charged $\pm {1\over 2}$ liptons transform
as doublets under $SU(2)'$ and are expected to
be confined into unstable integrally charged bound states.
These bound states are non-relativistic since the mass of the
lipton ( $> O$(100 GeV)) is expected to be much larger than
$\Lambda_{SU(2)'}$ ( $\le \Lambda_{QCD}$ ). The liptons also
participate in electroweak interactions so they could be
produced at colliders. Likewise the liptonic bound states can decay
into ordinary particles via electroweak interactions.
\vglue 0.6cm
{\elevenbf\noindent 3. Cosmological Implications}
\vglue 0.2cm
{\elevenit\noindent 3.1. Domain Walls} \cite{dw}
\vglue 0.1cm
\baselineskip=14pt
\elevenrm
Spontaneously broken discrete symmetries
can lead to cosmologically unacceptable domain walls. The
domain wall problem in the q-$\ell$ model can be avoided
as follows:
(i) Embed the discrete q-$\ell$ symmetry into a continuous
one. The resulting chain of symmetry breakings will produce
a string-wall network in which the walls become unstable.
(ii) There exists a parameter space for the Higgs potential
such that discrete q-$\ell$ symmetry restoration cannot occur at high
temperatures. This allows one to consistently arrange,
as an initial condition of the Big Bang, for all the vacua of
causally disconnected regions to be the same.
(iii) It could be possible that an inflationary period occurs
after the discrete q-$\ell$ symmetry phase transition during
the course of the evolution of the early universe.
\vglue 0.2cm
{\elevenit\noindent 3.2. $SU(2)'$ Glueballs} \cite{flvglue}
\vglue 0.1cm
If the $SU(2)'$ glueball has a mass around 1 GeV
then its lifetime is constrained to be less than 1 sec for
compatibility with standard Big Bang nucleosynthesis.
If the mass of the glueball is around 1 keV then its existence
will not interfere with nucleosynthesis and is a possible
candidate for Dark Matter.
\newpage
\vglue 0.5cm
{\elevenbf\noindent 4. References \hfil}
\vglue 0.2cm

\end{document}